\documentstyle[12pt]{article}

\begin{document}

\title{$U(3)_{L}\times U(3)_{R}$ Chiral Theory of Mesons}
\author{Bing An Li\\
Department of Physics and Astronomy, University of Kentucky\\
Lexington, KY 40506, USA}

\maketitle

\begin{abstract}
The chiral theory of mesons of two flavors have been extended to
mesons containing strange flavor. Two new mass relations between
vector and axial-vector mesons have been obtained. In
chiral limit, the physical
processes of normal parity and abnormal parity have been
studied. Due to the universality of the coupling of this theory
many interesting results have been obtained. In the chiral limit
theoretical results are in reasonable agreement with data.

\end{abstract}

\newpage
\newcommand{\ssp} {\partial \hspace{-.09in}/}
\newcommand{\sspp} {p \hspace{-.07in}/}
\newcommand{\pa} {\partial}
\newcommand{\ssv} {v \hspace{-.09in}/}
\newcommand{\ssD} {D \hspace{-.10in}/}
\newcommand{\ssa} {a \hspace{-.09in}/}
\newcommand{\ssA} {A \hspace{-.09in}/}
\newcommand{\ssW} {W \hspace{-.12in}/}
\newcommand{\da} {\dagger}
\newcommand{\ssun} {\underline{p} \hspace{-.09in}/}
\newcommand{\ssq} {q \hspace{-.09in}/}

\newpage
Chiral symmetry is
one of the most important features revealed from quantum chromodynamics
(QCD). Based on chiral symmetry and minimum coupling,
a meson theory of two flavors has been proposed[1] and the theoretical
results are in good agreement with the phenomenology of pseudoscalar,
vector, and axial-vector mesons made of $u$ and $d$ quarks.
In this paper we generalize the study of ref.[1] to $K$, $\eta$,
$\eta'$, $K^{*}(892)$, $\phi$, $K_{1}(1400)$, and $f_{1}(1510)$
mesons containing the third flavor-strange quark.
The paper is organized as follows.
1) the formalism of the theory;
2) new mass relations between vector and axial-vector mesons;
3) Vector meson dominance(VMD) and kion-form factors;
4) Decays of $\tau$ lepton;
5)Decays of $\phi$, $K^{*}$, $K_{1}(1400)$, $f_{1}(1510)$
and $\eta'$ mesons;
6) Decays of $K^{*}(892)\rightarrow K\pi\pi$;
7) Electromagnetic decays of mesons;
8) Summary of the results; 9) Conclusion.\\

{\large\bf The formalism of $U(3)_{L}\times U(3)_{R}$ chiral
theory of mesons}\\
Using $U(3)_{L}\times U(3)_{R}$ chiral symmetry and the minimum
coupling principle, the lagrangian of quarks of three flavors
and other fields
has been constructed as
\begin{eqnarray}
{\cal L}=\bar{\psi}(x)(i\ssp+\ssv+e_{0}Q\ssA+\ssa\gamma_{5}
-mu(x))\psi(x)
+{1\over 2}m^{2}_{1}(\rho^{\mu}_{i}\rho_{\mu i}+
\omega^{\mu}\omega_{\mu}+a^{\mu}_{i}a_{\mu i}+f^{\mu}f_{\mu})
\nonumber \\
+{1\over 2}m^{2}_{2}(K^{*a}_{\mu}\bar{K}^
{*a\mu}+K_{1}^{\mu}K_{1\mu})
+{1\over 2}m^{2}_{3}(\phi_{\mu}\phi^{\mu}+f_{s}^{\mu}f_{s\mu})
\nonumber \\
+\bar{\psi}(x)_{L}\ssW\psi(x)_{L}
+{\cal L}_{EM}+{\cal L}_{W}+{\cal L}_{lepton}
\end{eqnarray}
where \(a_{\mu}=\tau_{i}a^{i}_{\mu}+\lambda_{a}K^{a}_{1\mu}
+({2\over 3}+{1\over \sqrt{3}}\lambda_{8})
f_{\mu}+({1\over 3}-{1\over \sqrt{3}}\lambda_{8})
f_{s\mu}\)(\(i=1,2,3\) and \(a=4,5,6,7\)),
\(v_{\mu}=\tau_{i}
\rho^{i}_{\mu}+\lambda_{a}K^{*}_{\mu}+
({2\over 3}+{1\over \sqrt{3}}\lambda_{8})
\omega_{\mu}+({1\over 3}-{1\over \sqrt{3}}\lambda_{8})
\phi_{\mu}\), $A_{\mu}$ is photon field,
Q is the electric charge operator
of $u$, $d$, and $s$ quarks,
$W^{i}_{\mu}$ is W boson, and \(u=expi\{\gamma_{5}(\tau_{i}\pi_{i}+
\lambda_{a}K^{a}+\eta+
\eta')\}\), $m$ is a parameter. In eq.(1) $u$ can be written as
\begin{equation}
u={1\over 2}(1+\gamma_{5})U+{1\over 2}(1+\gamma_{5})U^{\da},
\end{equation}
where \(U=expi(\tau_{i}\pi_{i}+\lambda_{a}K^{a}+\eta+\eta')\).
In eq.(1) $\psi$ are $u$, $d$, and $s$ quark fields which carry
colors and other quantum numbers of quark. All other fields are
colorless. The physical fields related to $a_{\mu}$ and $v_{\mu}$
will be defined below.
As mentioned in ref.[1], in $QCD$ the boson fields $v_{\mu}$,
$a_{\mu}$, and pseudoscalars
are not fundamental fields and they should be bound state
solutions of $QCD$. Therefore, in eq.(1) there are no
kinetic terms for those fields and the kinetic terms will be
generated from quark loops(see below). As a matter of fact,
the relationship between boson fields and quark fields
can be found from lagrangian(1). Taking $\rho^{i}_{\mu}$ and
$a^{i}_{\mu}$ fields as examples. Using the least action principle
\[\frac{\delta {\cal L}}{\delta \rho^{i}_{\mu}}=0,\;\;\;
\frac{\delta {\cal L}}{\delta a^{i}_{\mu}}=0,\]
we obtain following relationships
\[\rho^{i}_{\mu}=-{1\over m^{2}_{1}}\bar{\psi}\tau_{i}\gamma_{\mu}
\psi,\;\;\;
a^{i}_{\mu}=-{1\over m^{2}_{1}}\bar{\psi}\tau_{i}\gamma_{\mu}\gamma_
{5}\psi.\]
Substituting these relations into eq.(1), except the term $-m\bar{\psi}
u\psi$ the hadronic part of the lagrangian(1) becomes Numbu-Jona-
Lasinio model[2]. The introduction of the pseudoscalar fields into
lagrangian(1) is based on nonlinear $\sigma$ model. u of eq.(1) is a
series of pseudoscalar fields. In principle, the relationships between
pseudoscalar fields and quark fields should be found by the least
action principle, but they are not as simple as the relations shown
above. This is the difference between present theory and Numbu-Jona-
Lasinio model.

It is the same with ref.[1] that
using method of path integral to integrating out
the quark fields, the effective lagrangian of mesons
are obtained.
To the fourth order in covariant derivatives in Minkofsky space,
the real part of the effective lagrangian
describing the physical processes of normal parity
takes following form
\begin{eqnarray}
\lefteqn{{\cal L}_{RE}=\frac{N_{c}}{(4\pi)^{2}}m^{2}{D\over 4}\Gamma
(2-{D\over 2})TrD_{\mu}UD^{\mu}U^{\da}}\nonumber \\
 & &-{1\over 3}\frac{N_{c}}{(4\pi
)^{2}}{D\over 4}\Gamma(2-{D\over 2})Tr\{
v_{\mu\nu}v^{\mu\nu}+a_{\mu\nu}a^{\mu\nu}\}\nonumber \\
 & &+{i\over 2}\frac{N_{c}}{(4\pi)^{2}}Tr\{D_{\mu}UD_{\nu}U^{\da}+
D_{\mu}U^{\da}D_{\nu}U\}v^{\nu\mu}\nonumber \\
 & &+{i\over 2}\frac{N_{c}}{(4\pi)^{2}}Tr\{D_{\mu}U^{\da}D_{\nu}U-D_{\mu}
UD_{\nu}U^{\da}\}a^{\nu\mu} \nonumber \\
 & &+\frac{N_{c}}{6(4\pi)^{2}}TrD_{\mu}D_{\nu}UD^{\mu}D^{\nu}U^{\da}
\nonumber \\
 & &-\frac{N_{c}}{12(4\pi)^{2}}Tr\{
D_{\mu}UD^{\mu}U^{\da}D_{\nu}UD^{\nu}U^{\da}
+D_{\mu}U^{\da}D^{\mu}UD_{\nu}U^{\da}D^{\nu}U
-D_{\mu}UD_{\nu}U^{\da}D^{\mu}UD^{\nu}U^{\da}\} \nonumber \\
 & &+{1\over 2}m^{2}_{1}(\rho^{\mu}_{i}\rho_{\mu i}+
\omega^{\mu}\omega_{\mu}+a^{\mu}_{i}a_{\mu i}+f^{\mu}f_{\mu})
\nonumber \\
 & &+{1\over 2}m^{2}_{2}(K^{*a}_{\mu}\bar{K}^
{*a\mu}+K_{1}^{\mu}K_{1\mu})
+{1\over 2}m^{2}_{3}(\phi_{\mu}\phi^{\mu}+f_{s}^{\mu}f_{s\mu}),
\end{eqnarray}
where
\begin{eqnarray*}
D_{\mu}U=\partial_{\mu}U-i[v_{\mu}, U]+i\{a_{\mu}, U\},\\
D_{\mu}U^{\da}=\partial_{\mu}U^{\da}-i[v_{\mu}, U^{\da}]-
i\{a_{\mu}, U^{\da}\},\\
v_{\mu\nu}=\partial_{\mu}v_{\nu}-\partial_{\nu}v_{\mu}
-i[v_{\mu}, v_{\nu}]-i[a_{\mu}, a_{\nu}],\\
a_{\mu\nu}=\partial_{\mu}a_{\nu}-\partial_{\nu}a_{\mu}
-i[a_{\mu}, v_{\nu}]-i[v_{\mu}, a_{\nu}],\\
D_{\nu}D_{\mu}U=\partial_{\nu}(D_{\mu}U)-i[v_{\nu}, D_{\mu}U]
+i\{a_{\nu}, D_{\mu}U\},\\
D_{\nu}D_{\mu}U^{\da}=\partial_{\nu}(D_{\mu}U^{\da})
-i[v_{\nu}, D_{\mu}U^{\da}]
-i\{a_{\nu}, D_{\mu}U^{\da}\}.
\end{eqnarray*}

Following ref.[1],
the effective lagrangian describing the physical processes with
abnormal party will be evaluated in terms of the quark propergators.

In this paper except the kion form factor $f_{-}(q^{2})$, all studies
have been done in the chiral limit.
In chiral limit, the following definitions[1] are held in this paper
\begin{eqnarray}
\frac{F^{2}}{16}=\frac{N_{c}}{(4\pi)^{2}}m^{2}\frac{D}{4}
\Gamma(2-{D\over 2}),\\
g^{2}={8\over 3}\frac{N_{c}}{(4\pi)^{2}}{D\over 4}
\Gamma(2-{D\over 2})={1\over 6}{F^{2}\over m^{2}}.
\end{eqnarray}
According to the arguments of ref.[1],
the physical meson fields have been defined as
\begin{eqnarray}
\rho\rightarrow {1\over g}\rho,\;\;\;
K^{*}\rightarrow {1\over g}K^{*},\;\;\;
\omega\rightarrow {1\over g}\omega,\;\;\;
\phi\rightarrow {\sqrt{2}\over g}\phi, \nonumber \\
a^{i}_{\mu}\rightarrow\frac{1}{g(1-{1\over 2\pi^{2}g^{2}})^
{{1\over 2}}}a^{i}_{\mu}
-{c\over g}\partial_{\mu}\pi^{i},\;\;\;
f_{\mu}\rightarrow\frac{1}{g(1-{1\over 2\pi^{2}g^{2}})^
{{1\over 2}}}f_{\mu}
-{c\over g}\partial_{\mu}\eta_{0},\nonumber \\
K_{1\mu}\rightarrow\frac{1}{g(1-{1\over 2\pi^{2}g^{2}})^
{{1\over 2}}}K_{1\mu}-
{c\over g}\partial_{\mu}K,\;\;\;
f_{s\mu}\rightarrow\frac{\sqrt{2}}{g(1-{1\over 2\pi^{2}g^{2}})^
{{1\over 2}}}f_{s\mu}
-{c\over g}\partial_{\mu}\eta_{s},\nonumber \\
\pi\rightarrow {2\over f_{\pi}}\pi,\;\;\;
K\rightarrow {2\over f_{K}}K, \;\;\;
\eta\rightarrow {2\over f_{\eta}}\eta, \;\;\;
\eta'\rightarrow {2\over f_{\eta'}}\eta',
\end{eqnarray}
where \(\eta_{0}=({1\over \sqrt{3}}cos\theta-\sqrt{{2\over 3}}sin
\theta)\eta+({1\over \sqrt{3}}sin\theta+\sqrt{{2\over 3}}cos\theta)
\eta'\) and \(\eta_{s}=(-{2\over \sqrt{3}}cos\theta-\sqrt{2\over 3}
sin\theta)\eta+(-{2\over \sqrt{3}}sin\theta+\sqrt{2\over 3}cos\theta)
\eta'\), $\theta$ is the mixing angle of $\eta$ and $\eta'$.
In chiral limit we take \(f_{\pi}=f_{K}=f_{\eta}=f_{\eta'}\).
In chiral limit following two equations of ref.[1] are held
in the case of three flavors
\begin{eqnarray}
c=\frac{f^{2}_{\pi}}{2gm^{2}_{\rho}},\\
{F^{2}\over f^{2}_{\pi}}(1-{2c\over g})=1.
\end{eqnarray}
Following ref.[1] we have
\begin{equation}
g=0.35.
\end{equation}
Use the substitutions(6), the physical masses of vector
masons are defined as
\begin{equation}
m^{2}_{\rho}=m^{2}_{\omega}={1\over g^{2}}m^{2}_{1},\;\;\;
m^{2}_{K^{*}}={1\over g^{2}}m^{2}_{2},\;\;\;
m^{2}_{\phi}={2\over g^{2}}m^{2}_{3}.
\end{equation}

{\large\bf New mass formulas of vector mesons and its chiral
partners}\\

In ref.[1] two mass relations of $a_{1}$ $\rho$ and $f_{1}(1285)$
$\omega$ have been obtained
\begin{eqnarray}
(1-\frac{1}{2\pi^{2}g^{2}})m^{2}_{a}={F^{2}\over g^{2}}+m^{2}_{\rho},
\nonumber \\
(1-\frac{1}{2\pi^{2}g^{2}})m^{2}_{f}={F^{2}\over g^{2}}+m^{2}_{\omega}.
\end{eqnarray}
By the same reasons obtaining eqs.(11), we obtain
other two mass formulas
\begin{eqnarray}
(1-\frac{1}{2\pi^{2}g^{2}})m^{2}_{K_{1}}={F^{2}\over g^{2}}+
m^{2}_{K^{*}},\nonumber \\
(1-\frac{1}{2\pi^{2}g^{2}})m^{2}_{f_{1}(1510)}={F^{2}\over g^{2}}+
m^{2}_{\phi}.
\end{eqnarray}
If input $m_{a}$, $m_{\rho}$, and $f_{\pi}$, we obtain
\begin{equation}
m_{f_{1}(1285)}=1.27GeV,\;\;m_{K_{1}(1400)}=1.38GeV,\;\;
m_{f_{1}(1510)}=1.51GeV.
\end{equation}
The deviations from data are about $1\%$. In Table I, the masses of
these mesons are obtained by taking \(g=0.35\).

Weinberg's first sum rule[3] is
\begin{equation}
\frac{g^{2}_{\rho}}{m^{2}_{\rho}}-\frac{g^{2}_{a}}{m^{2}_{a}}=
{1\over 4}f^{2}_{\pi},
\end{equation}
where $g_{a}$ and $g_{\rho}$ are defined in ref.[1].
In order to compare with this sum rule, the four mass formulas(11,12)
can be rewritten as
\begin{eqnarray}
\frac{m^{2}_{a}}{g^{2}_{a}}-\frac{m^{2}_{\rho}}{g^{2}_{\rho}}=
\frac{f^{2}_{\pi}m^{4}_{\rho}}{g^{2}_{\rho}(4g^{2}_{\rho}-f^{2}_{\pi}
m^{2}_{\rho})},\nonumber \\
\frac{m^{2}_{f_{1}(1285)}}{g^{2}_{a}}-\frac{m^{2}_{\omega}}
{g^{2}_{\rho}}=
\frac{f^{2}_{\pi}m^{4}_{\rho}}{g^{2}_{\rho}(4g^{2}_{\rho}-f^{2}_{\pi}
m^{2}_{\rho})},\nonumber \\
\frac{m^{2}_{K_{1}(1400)}}{g^{2}_{a}}-\frac{m^{2}_{K^{*}}}
{g^{2}_{\rho}}=
\frac{f^{2}_{\pi}m^{4}_{\rho}}{g^{2}_{\rho}(4g^{2}_{\rho}-f^{2}_{\pi}
m^{2}_{\rho})},\nonumber \\
\frac{m^{2}_{f_{1}(1510)}}{g^{2}_{a}}-\frac{m^{2}_{\phi}}
{g^{2}_{\rho}}=
\frac{f^{2}_{\pi}m^{4}_{\rho}}{g^{2}_{\rho}(4g^{2}_{\rho}-f^{2}_{\pi}
m^{2}_{\rho})}.
\end{eqnarray}

{\large\bf Vector meson dominance(VMD) and kion form factors} \\
Vector meson dominance(VMD) has been revealed from this theory[1].
Besides the $\rho$ and $\omega$ dominance
\begin{eqnarray}
{e\over f_{\rho}}\{-{1\over 2}F^{\mu\nu}(\partial_{\mu}\rho^{0}_
{\nu}-\partial_{\nu}\rho^{0}_{\mu})+A^{\mu}j^{0}_{\mu}\},\nonumber \\
{e\over f_{\omega}}\{-{1\over 2}F^{\mu\nu}(\partial_{\mu}\omega_
{\nu}-\partial_{\nu}\omega_{\mu})+A^{\mu}j^{\omega}_{\mu}\},
\nonumber \\
{1\over f_{\rho}}={1\over 2}g,\;\;\;{1\over f_{\omega}}=
{1\over 6}g,
\end{eqnarray}
the $\phi$ dominance has been derived from
eq.(1)
\begin{equation}
{e\over f_{\phi}}\{-{1\over 2}F^{\mu\nu}(\partial_{\mu}\phi_
{\nu}-\partial_{\nu}\phi_{\mu})+A^{\mu}j^{\phi}_{\mu}\},\nonumber \\
f_{\phi}=-{1\over 3\sqrt{2}}g,
\end{equation}
where $j^{\phi}_{\mu}$ is the current that $\phi$ meson couples to.

The electric kion form factor can be studied by VMD.
The effective lagrangian
of $KK\gamma$ consists of two parts: kions couple to photon directly
and kions couple to vector mesons then the vector mesons couple
to photon. In chiral limit, the
couplings between kions and $\rho$, $\omega$, and $\phi$ mesons can
be found from eq.(3)
\begin{eqnarray}
{\cal L}_{K\bar{K}v}=-{2\sqrt{2}i\over g}
\phi_{\mu}(K^{+}\partial_{\mu}K^{-}+K^{0}\partial_{\mu}\bar{K}^{0}),
\nonumber \\
+{2i\over g}
\omega_{\mu}(K^{+}\partial_{\mu}K^{-}+K^{0}\partial_{\mu}\bar{K}^{0}),
\nonumber \\
+{2i\over g}
\rho_{\mu}(K^{+}\partial_{\mu}K^{-}-K^{0}\partial_{\mu}\bar{K}^{0}).
\end{eqnarray}
In eq.(18), the eq.(8) has been used. Using the substitutions
\begin{equation}
\rho_{\mu}\rightarrow {e\over f_{\rho}}A_{\mu},\;\;\;
\omega_{\mu}\rightarrow {e\over f_{\omega}}A_{\mu},\;\;\;
\phi_{\mu}\rightarrow {e\over f_{\phi}}A_{\mu},
\end{equation}
in eqs.(18) the direct couplings of $KK\gamma$ can be obtained.
Using the couplings $-{1\over 2}{e\over f_{v}}F_{\mu\nu}(\partial
_{\mu}v_{\nu}-\partial_{\nu}v_{\mu})$ (\(v=\rho, \omega, \phi\))
of eqs.(16,17) and eqs.(18), the indirect couplings of $KK\gamma$ can
be obtained. Adding these two couplings together, the electric
form factor of charged kion has been obtained
\begin{eqnarray}
F_{K^{+}}(q^{2})= {1\over 3}\frac{m^{2}_{\phi}}{m^{2}_{\phi}-
q^{2}}+{1\over 6}\frac{m^{2}_{\omega}}{m^{2}_{\omega}-q^{2}}
+{1\over 2}\frac{m^{2}_{\rho}}{m^{2}_{\rho}-q^{2}}.
\end{eqnarray}
Because of eq.(8) this form factor is normalized to be one at
\(q^{2}=0\) and the radius is determined to be
\begin{equation}
<r^{2}>={2\over m^{2}_{\phi}}+{1\over m^{2}_{\omega}}+
{3\over m^{2}_{\rho}}=0.33fm^{2}.
\end{equation}
The theoretical result agrees with data[4](see Table I). In the
same way, the electric form factor of neutral kion has been obtained
\begin{eqnarray}
F_{K^{0}}(q^{2})= {1\over 3}\frac{m^{2}_{\phi}}{m^{2}_{\phi}-
q^{2}}+{1\over 6}\frac{m^{2}_{\omega}}{m^{2}_{\omega}-q^{2}}
-{1\over 2}\frac{m^{2}_{\rho}}{m^{2}_{\rho}-q^{2}},\nonumber \\
\frac{\partial F_{K^{0}}(q^{2})}{\partial q^{2}}|_{q^{2}=0}=
{1\over 3m^{2}_{\phi}}+{1\over 6m^{2}_{\omega}}-{1\over 2
m^{2}_{\rho}}=-0.25 GeV^{-2}.
\end{eqnarray}

The VMD can be applied to study the form factors of $K\rightarrow\pi
l\nu$. Let's study $K^{+}\rightarrow \pi^{0}l\nu$ first.
The vertex of $K^{*}(892)K\pi^{0}$ has normal parity and
can be derived from eq.(3)
\begin{eqnarray}
{\cal L}_{K^{*}K\pi^{0}}={i\over g}
\{(\pi^{0}\partial^{\mu}K^{-}-\partial_{\mu}\pi^{0}K^{-})K^{+\mu}
+(\pi^{0}\partial_{\mu}K^{+}-\partial_{\mu}\pi^{0}K^{+})K^{-\mu}\}
\nonumber \\
+\frac{4i}{gf^{2}_{\pi}}\{c^{2}-\frac{2N_{c}}{(4\pi)^{2}}(1-
{2c\over g})^{2}\}\partial^{\nu}\pi^{0}\{(\partial_{\mu}K^{-}_{\nu}
-\partial_{\nu}K^{-}_{\mu})\partial^{\mu}K^{-}\}\nonumber \\
-\frac{i}{8\pi^{2}g}(1-{2c\over g})^{2}\{(\partial_{\nu}\pi^{0}
\partial^{\mu\nu}K^{-}-\partial_{\nu}K^{-}\partial_{\mu\nu}\pi^{0})
K^{+\mu}+\partial_{\nu}\pi^{0}\partial^{\mu\nu}K^{+}-
\partial^{\mu\nu}\pi^{0}\partial_{\nu}K^{+})K^{-}_{\mu}\}
\end{eqnarray}
In the decays of $K\rightarrow\pi l\nu$, there are direct coupling
$K\pi W$ and indirect couplings $K\pi K^{*}$ and $K^{*}W$. Using the
substitution obtained from eq.(1)
\begin{equation}
K_{\mu}\rightarrow {g_{w}\over 4}gW_{\mu}sin\theta_{c}
\end{equation}
in eq.(23), the direct coupling can be obtained. It is similar to
VMD, the coupling between $K^{*}$ and W-boson has been obtained
from eq.(3)
\begin{equation}
{\cal L}_{i}=-{1\over 4}gg_{w}\{(\partial_{\mu}K^{+}_{\nu}-
\partial_{\nu}K^{+}_{\mu})\partial^{\mu}W^{-\nu}
+(\partial_{\mu}K^{-}_{\nu}-
\partial_{\nu}K^{-}_{\mu})\partial^{\mu}W^{+\nu}\}
\end{equation}
{}From eqs.(23,24,25) the indirect coupling has been obtained.
Adding the direct
and indirect couplings together, the two form factors of $K^{+}
\rightarrow \pi^{0}l\nu$ have been obtained
\begin{eqnarray}
f_{+}(q^{2})={1\over \sqrt{2}}\frac{m^{2}_{K^{*}}}{m^{2}_{K^{*}}
-q^{2}},\;\;\;f_{-}(q^{2})=-{1\over \sqrt{2}}{1\over m^{2}_{K^{*}}
-q^{2}}(m^{2}_{K^{+}}-m^{2}_{\pi^{0}}),\nonumber \\
\lambda_{+}=-\lambda_{-}=\frac{m^{2}_{\pi^{0}}}{m^{2}_{K^{*}}}=0.0239,
\nonumber \\
\xi=\frac{f_{-}}{f_{+}}=-\frac{m^{2}_{K^{+}}-m^{2}_{\pi^{0}}}{m^{2}
_{K^{+*}}}=-0.284.
\end{eqnarray}
For $K^{0}_{l3}$ we have obtained following quantities in the same
way with $K^{+}\rightarrow\pi^{0}l\nu$
\begin{eqnarray}
f_{+}(q^{2})={1\over \sqrt{2}}\frac{m^{2}_{K^{*}}}{m^{2}_{K^{*}}
-q^{2}},\;\;\;f_{-}(q^{2})=-{1\over \sqrt{2}}{1\over m^{2}_{K^{*}}
-q^{2}}(m^{2}_{K^{0}}-m^{2}_{\pi^{+}}),\nonumber \\
\lambda_{+}=-\lambda_{-}=\frac{m^{2}_{\pi^{+}}}{m^{2}_{K^{*}}}=0.0245,
\nonumber \\
\xi=\frac{f_{-}}{f_{+}}=-\frac{m^{2}_{K^{0}}-m^{2}_{\pi^{+}}}{m^{2}
_{K^{+*}}}=-0.287.
\end{eqnarray}
In eqs.(26,27), the leading terms of chiral perturbation have been kept.
The decay widths have been computed
\begin{equation}
\Gamma(K^{+}_{e3})=0.233\times 10^{-17} GeV,\;\;\;
\Gamma(K^{0}_{e3})=0.483\times 10^{-17} GeV.
\end{equation}
The comparisons with experimental data are shown in Table I.

{\large\bf Decays of $\tau\rightarrow K^{*}(892)\nu$ and
$\tau\rightarrow K_{1}(1400)\nu$}\\
It is similar to the calculation of
$\tau\rightarrow \rho\nu$ and $\tau\rightarrow
a_{1}\nu$[1], we obtain
\begin{eqnarray}
\lefteqn{\Gamma(\tau\rightarrow K^{*}(892)\nu)=\frac{G^{2}}{32\pi}
sin^{2}\theta_{c}
g^{2}m^{2}_{K^{*}}m^{3}_{\tau}(1-{m^{2}_{K^{*}}\over m^{2}_{\tau}})
^{2}(1+2{m^{2}_{K^{*}}\over m^{2}_{\tau}})=0.326\times 10^{-13} GeV,}
\nonumber \\
&&B(\tau\rightarrow K^{*}(892)\nu)=1.46\%,\nonumber \\
&&\Gamma(\tau\rightarrow K_{1}\nu)=\frac{G^{2}}{32\pi}
sin^{2}\theta_{c}
g^{2}(1-{1\over 2\pi^{2}g^{2}})^{-1}
\frac{m^{4}_{K^{*}}}{m^{2}_{K_{1}}}m^{3}_{\tau}
(1-{m^{2}_{K^{*}}\over m^{2}_{\tau}})
^{2}(1+2{m^{2}_{K^{*}}\over m^{2}_{\tau}})=0.831\times 10^{-14} GeV,
\nonumber \\
&&B(\tau\rightarrow K^{*}(1400)\nu)=0.373\%.
\end{eqnarray}

{\large\bf Decays of $\phi$, $K^{*}(892)$, $K_{1}(1400)$, and
$f_{1}(1510)$ mesons}\\
In this theory, the vertices of $\phi K\bar{K}$, $K^{*}K\pi$,
$K_{1}K^{*}\pi$, $K_{1}K\rho$, $K_{1}K\omega$, and
$f_{1}(1510)K^{*}\bar{K}$ contain
even numbers of $\gamma_{5}$, therefore, they
are the processes with normal parity and the vertices of these
processes can be derived from eq.(3). In this section the calculation
of the decay widths of theses processes have been provided.

{\bf Decays of $\phi\rightarrow K\bar{K}$} \\
In chiral limit, the vertex of this process has been found from eq.(3)
\begin{equation}
{\cal L}_{\phi K\bar{K}}=\frac{i2\sqrt{2}}{g}\phi_{\mu}(K^{+}
\partial^{\mu}K^{-}+K^{0}\partial^{\mu}\bar{K}^{0}).
\end{equation}
In deriving eq.(30), eq.(8) have been used. The numerical results of the
decays are
\begin{eqnarray}
\Gamma(\phi\rightarrow K^{0}\bar{K}^{0})=1.11 MeV,\;\;\;
\Gamma(\phi\rightarrow K^{+}K^{-})=1.7 MeV,\;\;
\frac{\Gamma(\phi\rightarrow K^{0}\bar{K}^{0})}{\Gamma(\phi\rightarrow
K^{+}K^{-})}=0.66.
\end{eqnarray}
{\bf Decays of $K^{*}(892)\rightarrow K\pi$}\\
In chiral limit and using eq.(3), the vertex of this process has been
obtained
\begin{eqnarray}
{\cal L}_{K^{*}K\pi}=\frac{2i}{g}\{\sqrt{2}\pi^{+}(K^{-}_{\mu}\partial
^{\mu}K^{0}-K^{0}_{\mu}\partial^{\mu}K^{-})+\sqrt{2}\pi^{-}(\bar{K}
^{0}_{\mu}\partial^{\mu}K^{+}-K^{+}_{\mu}\partial^{\mu}\bar{K}^{0})
\nonumber \\
+\pi^{0}(K^{-}_{\mu}\partial^{\mu}K^{+}-K^{+}_{\mu}\partial^{\mu}K^{-}
-\bar{K}^{0}_{\mu}\partial^{\mu}K^{0}
+K^{0}_{\mu}\partial^{\mu}\bar{\mu}K^{0})\}
\end{eqnarray}
The numerical results of the decay widths are
\begin{equation}
\Gamma(K^{*}\rightarrow K^{0}\pi^{+})=25.4 MeV,\;\;\;\Gamma(K^{*}
\rightarrow K^{+}\pi^{0})=14.0 MeV,\;\;\; \Gamma_{tot}=39.4 MeV.
\end{equation}

{\bf Decays of $K_{1}(1400)$}\\
In chiral limit the vertex of $K_{1}\rightarrow K^{*}\pi$ has been
derived from eq.(3)
\begin{eqnarray}
{\cal L}_{K_{1}K^{*}\pi}=f_{abi}\pi^{i}\{AK^{*a}_{1\mu}K^{b\mu}
+Bp_{\pi}^{\mu}p_{\pi}^{\nu}K^{a}_{1\mu}K^{*b}_{\nu}\},
\nonumber \\
A={2\over f_{\pi}}(1-{1\over 2\pi^{2}g^{2}})^{-{1\over 2}}
(m^{2}_{K_{1}}-m^{2}_{K^{*}})(1-{2c\over g})
(1-{3\over 4\pi^{2}g^{2}}),
\nonumber \\
B=-{2\over f_{\pi}}(1-{1\over 2\pi^{2}g^{2}})^{-{1\over 2}}
{1\over 2\pi^{2}g^{2}}(1-{2c\over g}),
\end{eqnarray}
where $m^{2}_{K_{1}}$ is determined by eq.(12). In the expression of A
(34), eq.(12) has been applied.
The numerical results are
\begin{equation}
\Gamma(K_{1}\rightarrow K^{*+}\pi^{0})=42.MeV,\;\;\;\Gamma(K_{1}
\rightarrow K^{*0}\pi^{+})=2\Gamma(K_{1}\rightarrow K^{*+}\pi^{0}),
\;\;\;\Gamma_{tot}=126.MeV.
\end{equation}
The vertex of $K_{1}K\rho$ can be found from eq.(3) and it is just
the formula obtained by
using following formula as the A in eq.(34)
\begin{equation}
A={2\over f_{\pi}}(1-{1\over 2\pi^{2}g^{2}})^{-{1\over 2}}\{
m^{2}_{K_{1}}-m^{2}_{K^{*}}-(m^{2}_{K_{1}}-m^{2}_{\rho})[{2c\over g}
+\frac{3}{4\pi^{2}g^{2}}(1-{2c\over g})]\}.
\end{equation}
The decay width of $K_{1}\rightarrow K\rho$ has been calculated
\begin{equation}
\Gamma(K_{1}\rightarrow K\rho)=19.3 MeV,\;\;\;
Branch\; ratio=11.1(1\pm 0.075)\%
\end{equation}
In the same way, if ignore the mass difference of $\rho$ and $\omega$
mesons we obtain
\begin{equation}
\Gamma(K_{1}\rightarrow K\omega)={1\over 3}\Gamma(K_{1}\rightarrow K
\rho).
\end{equation}
The numerical results are
\begin{equation}
\Gamma(K_{1}\rightarrow K\omega)=4.12 MeV,\;\;\;
Branch\; ratio=2.4\%.
\end{equation}
Comparing $\Gamma(K_{1}\rightarrow K\rho)$ and $\Gamma(K_{1}\rightarrow
K\omega)$ with $\Gamma(K_{1}\rightarrow K^{*}\pi)$, the formers are
much less than the later. Except the phase space, the differences
of the formulas for
these three processes are caused by
the masses of $\rho$, $\omega$ and $K^{*}$
in the amplitude A. The cancellations in A(eq.(36)) cause the
smallness of
$\Gamma(K_{1}\rightarrow K\rho)$ and
$\Gamma(K_{1}\rightarrow K\omega)$.

{\bf Decay of $K_{1}\rightarrow K\gamma$}\\
Using VMD(16), following vertex has been derived from the vertex
of $K_{1}K\rho$
\begin{equation}
{\cal L}_{K_{1}K\gamma}=-{i\over 2}e({1\over f_{\rho}}+
{1\over f_{\omega}}-{1\over f_{\phi}}){2\over f_{\pi}}(1-
{1\over 2\pi^{2}g^{2}})^{-{1\over 2}}\{m^{2}_{K_{1}}-m^{2}_{K^{*}}-
m^{2}_{K_{1}}[{2c\over g}+\frac{3}{4\pi^{2}g^{2}}(1-{2c\over g})]\}.
\end{equation}
The numerical result is
\begin{equation}
\Gamma(K_{1}\rightarrow K\gamma)=440keV.
\end{equation}

{\bf Decays of $f_{1}(1510)\rightarrow K^{*}(892)\bar{K}$}\\
{}From eq.(3), the decay amplitude has been obtained
\begin{eqnarray}
\lefteqn{<K^{+}(p_{1})K^{*-}(p_{2})|S|f_{1}(p)>=}\nonumber \\
&&-(2\pi)^{4}\delta^{4}(p-p_{1}-p_{2})\frac{1}{\sqrt{8m_{f}E_{K}
E_{K^{*}}}}e^{\lambda}_{\mu}(p)e^{\lambda'*}_{\nu}\{Ag^{\mu\nu}+
Bp^{\mu}_{1}p^{\nu}_{2}\},\nonumber \\
&&A={1\over f_{\pi}}(1-{1\over 2\pi^{2}g^{2}})^{-{1\over 2}}(
m^{2}_{f}-m^{2}_{K^{*}})(1-{2c\over g})(1-{3\over 4\pi^{2}g^{2}}),
\nonumber \\
&&B=-{1\over f_{\pi}}(1-{1\over 2\pi^{2}g^{2}})^{-{1\over 2}}\frac{1}
{2\pi^{2}g^{2}}(1-{2c\over g}).
\end{eqnarray}
There are four channels in this decays and the numerical results are
\begin{equation}
\Gamma(f_{1}\rightarrow K^{+}K^{*-})=5.48 MeV,\;\;\;
\Gamma_{tot}=21.9 MeV.
\end{equation}

{\bf Decays of $\eta'\rightarrow \eta\pi\pi$}\\
In this theory, the vertex of this process contains four $\gamma_{5}$.
Therefore, this is a process of normal parity and the vertex should
be derived from the eq.(3). It is well known that the masses of pion
and
$\eta$ are proportional to light quark masses[5], therefore,
in chiral
limit $m_{\pi}, m_{\eta}\rightarrow 0$. However, due to the $U(1)$
problem[6] $m^{2}_{\eta'}$ does not approach to zero in the limit
of chiral symmetry. Therefore, in chiral limit only the mass of
$\eta'$ meson has been kept in the amplitude of $\eta'\rightarrow
\eta\pi\pi$.
The calculation shows that in eq.(3) only the terms at the fourth order
in derivatives contribute to $\eta'\rightarrow \eta\pi\pi$.
Consequently, in the amplitude of this decay there is a factor of
${1\over (4\pi)^{2}}$. Therefore, this theory predicts that the width
of this decay is very narrow. The amplitude has been found
\begin{eqnarray}
\lefteqn{<\pi^{0}(k_{1})\pi^{0}(k_{2})\eta(p)|S|\eta'(p')>
=
i(2\pi)^{4}\delta^{4}(p'-p-k_{1}-k_{2})\frac{1}{\sqrt{16m_{\eta'}
E_{\eta}\omega_{1}\omega_{2}}}}\nonumber \\
&&{8\over f^{4}_{\pi}}\frac{2}{(4\pi)^{2}}
\{{1\over 2}(1-{2c\over g})^{4}(q^{4}_{1}+q^{4}_{2}+q^{4}_{3})+
(1-{2c\over g})[\frac{2c^{2}}{g^{2}}-(1-{2c\over g})^{2}]m^{4}_{\eta'}
\nonumber \\
&&+(1-{2c\over g})[{1\over 2}(1-{2c\over g})
+{4c^{3}\over g^{3}}]q^{2}_{3}m^{2}_{\eta'}
+{1\over 2}(1-{2c\over g})^{2}(1-{4c^{2}\over g^{2}})(q^{2}_{1}+
q^{2}_{2})m^{2}_{\eta'}\},
\end{eqnarray}
where \(q^{2}_{1}=(p'-k_{1})^{2}\), \(q^{2}_{2}=(p'-k_{2})^{2}\),
\(q^{2}_{3}=(p'-p)^{2}\).
The contribution of quark masses to the mass of $\eta'$ is about
0.376GeV,
therefore, in chiral limit \(m_{\eta'}=0.582GeV\). Using this value
we obtain
\begin{equation}
\Gamma(\eta'\rightarrow\eta\pi^{+}\pi^{-})=85.7keV,\;\;\;
\Gamma(\eta'\rightarrow\eta\pi^{0}\pi^{0})=48.6keV.
\end{equation}
In the range of $(0.958)^{2}\geq m^{2}_{\eta'}\geq 0$, we obtain
\begin{equation}
22.1keV\leq\Gamma(\eta'\rightarrow\eta\pi^{+}\pi^{-})\leq 145.2 keV
,\;\;\;
12.5keV\leq\Gamma(\eta'\rightarrow\eta\pi^{0}\pi^{0})\leq
82.4keV.
\end{equation}
Eq.(46) tells that, indeed, the decay widths are always small and
the data(see Table I) prefers a nonzero $m_{\eta'}$ in chiral limit.
This is consistent with the study of $U(1)$ problem in
$m_{\eta'}$[6]. Therefore, phenomenologically, in lagrangian(1) a mass
term of $\eta'$ should be added. The study of the $U(1)$ problem
could bring something new to present theory. However, this is not
the task of this paper.

{\large\bf Decays of $K^{*}(892)\rightarrow K\gamma$ and $K\pi\pi$}\\
The decays of $K^{*}\rightarrow K\gamma$ and $K\pi\pi$ have been
studied in ref.[7] by using gauging Wess-Zumino lagrangian. As
mentioned in ref.[1], the formalism obtained from this theory is the
same with the one in ref.[6]. However, in this theory the couplings
are universal and VMD is a result of present theory and not an input.
According to VMD, the decays of $K^{*}\rightarrow K\gamma$ are
associated with $K^{*}\rightarrow K v$. Therefore, the processes of
$K^{*}\rightarrow K\gamma$ have abnormal parity. The vertices of
$K^{*}Kv$ can be found from the calculation of ${1\over g}K^{*}_
{a\mu}<\bar{\psi}\lambda_{a}\gamma^{\mu}\psi>$, which is similar
with ${1\over g}\omega_{\mu}<\bar{\psi}\gamma^{\mu}\psi>$ in ref.[1],
\begin{equation}
{\cal L}_{K^{*}Kv}=-\frac{N_{c}}{2g^{2}\pi^{2}}{2\over f_{\pi}}
\varepsilon_
{\mu\nu\alpha\beta}d_{abc}K^{*}_{a\mu}\partial_{\nu}v^{c}_{\alpha}
\partial_{\beta}P^{b},
\end{equation}
where $P^{b}$ is a pseudoscalar meson and $v^{i}_{\alpha}$ is a
vector meson. From eq.(47) following vertices have been obtained
\begin{equation}
{\cal L}_{i}=-\frac{N_{c}}{2\pi^{2}g^{2}}{2\over f_{\pi}}
\varepsilon^{\mu\nu\alpha
\beta}K^{*+}_{\mu}\partial_{\beta}K^{+}\{{1\over 2}\partial_{\nu}
\rho^{0}_{\alpha}+{1\over 2}\partial_{\nu}\omega_{\alpha}
+{\sqrt{2}\over 2}\partial_{\nu}\phi_{\alpha}\}.
\end{equation}
Using VMD(eqs.(16,17)), we obtain
\begin{equation}
{\cal L}_{K^{+*}K^{+}\gamma}=-\frac{e}{4\pi^{2}g}{2\over f_{\pi}}
\varepsilon^{\mu\nu\alpha
\beta}K^{+}_{\mu}\partial_{\beta}K^{+}\partial_{\nu}A_{\alpha}.
\end{equation}
The the decay width has been computed
\begin{equation}
\Gamma(K^{+*}\rightarrow K^{+}\gamma)=43.5 keV.
\end{equation}
In the same way, it has been obtained
\begin{equation}
{\cal L}_{K^{0*}K^{0}\gamma}=\frac{e}{2\pi^{2}g}{2\over f_{\pi}}
\varepsilon^{\mu\nu\alpha
\beta}K^{0}_{\mu}\partial_{\beta}\bar{K}^{0}\partial_{\nu}A_{\alpha},
\end{equation}
and the decay width is
\begin{equation}
\Gamma(K^{0*}\rightarrow K^{0}\gamma)=175.4 keV.
\end{equation}

The experimental value of the branch ratio of $K^{*}\rightarrow
K\pi\pi$ is less than $7\times 10^{-4}$[4].
To understand so small
branch ratio is a crucial test for present theory. There are three
channels
\[K^{-*}\rightarrow K^{-}\pi^{0}\pi^{0}, K^{-}\pi^{+}\pi^{-},
\bar{K}^{0}\pi^{-}\pi^{0}.\]
The decay $K^{-*}\rightarrow K^{-}\pi^{0}\pi^{0}$ consists of
$K^{-*}\rightarrow K^{-*}\pi^{0}$ and $K^{-*}\rightarrow
K^{-}\pi^{0}$. The vertices have been obtained from eqs.(47,32)
\begin{eqnarray}
{\cal L}_{K^{-*}K^{+*}\pi^{0}}=-\frac{N_{c}}{\sqrt{2}\pi^{2}
g^{2}}{2\over f_{\pi}}\varepsilon^{\mu\nu\alpha\beta}
K^{-}_{\mu}\partial_{\nu}K^{+}_{\alpha}\partial_{\beta}\pi^{0},
\nonumber \\
{\cal L}_{K^{-*}K^{+}\pi^{0}}={2i\over g}K^{-}_{\mu}\partial^{\mu}
K^{+}\pi^{0}.
\end{eqnarray}
These two vertices lead to following amplitude of $K^{-*}\rightarrow
K^{-}\pi^{0}\pi^{0}$
\begin{equation}
{\cal M}=\frac{\sqrt{2}N_{c}}{\pi^{2}g^{3}}{2\over f_{\pi}}
\varepsilon^{\mu\nu\alpha\beta}\epsilon^{\lambda}_{\mu}(p')
p'_{\nu}k_{1\alpha}k_{2\beta}\{\frac{1}{(p'-k_{2})^{2}-m^{2}_{K^{*}}}
-\frac{1}{(p'-k_{1})^{2}-m^{2}_{K^{*}}}\},
\end{equation}
where $p'$, p, $k_{1}$, $k_{2}$ are momentum of $K^{*}$, K,
$\pi^{0}$, and $\pi^{0}$ respectively. It can be seen
that there is cancellation in eq.(54). This
cancellation has been obtained in ref.[7]. The width calculated is
\begin{equation}
\Gamma(K^{-*}\rightarrow K^{-}\pi^{0}\pi^{0})=0.214keV.
\end{equation}

The second channel $K^{-*}\rightarrow K^{-}\pi^{+}\pi^{-}$
consists of three processes: direct coupling $K^{-*}K^{+}\pi^{+}
\pi^{-}$ and indirect couplings:
$K^{-*}\rightarrow \bar{K}^{0*}\pi^{-}$ and $\bar{K}
^{0*}\rightarrow K^{-}\pi^{+}$, $K^{-*}\rightarrow K^{-}\rho^{0}$
and $\rho^{0}\rightarrow \pi^{+}\pi^{-}$.
The direct coupling has been found from ${1\over g}K^{a\mu}
<\bar{\psi}\lambda_{a}\gamma_{\mu}\psi>$ whose calculation is
similar with the one from which the direct coupling $\omega\pi\pi\pi$
has been found in ref.[1]
\begin{equation}
{\cal L}_{K^{*}K\pi\pi}=\frac{1}{4\pi^{2}g}({2\over f_{\pi}})^{3}
(1-{6c\over g}+
{6c^{2}\over g^{2}})\varepsilon^{\mu\nu\alpha\beta}K^{a}_{\mu}
\partial_{\nu}P^{b}\partial_{\alpha}P^{c}\partial_{\beta}P^{d}
d_{abe}f_{cde},
\end{equation}
where P stands for pseudoscalar field. Eq.(56) leads to
\begin{equation}
{\cal L}_{K^{-*}K^{+}\pi^{+}\pi^{-}}=-\frac{i}{2\sqrt{2}\pi^{2}g}
({2\over f_{\pi}})^{3}(1-{6c\over g}+
{6c^{2}\over g^{2}})\varepsilon^{\mu\nu\alpha\beta}K^{-}_{\mu}
\partial_{\nu}K^{+}\partial_{\alpha}\pi^{-}\partial_{\beta}
\pi^{+}.
\end{equation}
{}From eqs.(32,47) following vertices have been obtained
\begin{eqnarray}
\lefteqn{{\cal L}_{K^{-*}K^{0*}\pi^{+}}=
-\frac{N_{c}}{2\sqrt{2}\pi^{2}g^{2}}
{2\over f_{\pi}}\varepsilon^{\mu\nu\alpha\beta}K^{-}_{\mu}
\partial_{\nu}K^{0}_{\alpha}\partial_{\beta}\pi^{+},}\nonumber \\
&&{\cal L}_{\bar{K}^{0*}K^{+}\pi^{-}}=\frac{2\sqrt{2}i}{g}\pi^{-}
\bar{K}^{0}_{\mu}\partial^{\mu}K^{+},\nonumber \\
&&{\cal L}_{K^{-*}K^{+}\rho^{0}}=-\frac{N_{c}}{4\pi^{2}g^{2}}
{2\over f_{\pi}}\varepsilon^{\mu\nu\alpha\beta}K^{-}_{\mu}
\partial_{\beta}K^{+}\partial_{\nu}\rho^{0}_{\alpha},\nonumber \\
&&{\cal L}_{\rho^{0}\pi\pi}={2\over g}\epsilon_{3jk}\rho^{0}_{\mu}
\pi_{j}\partial^{\mu}\pi_{k},
\end{eqnarray}
where ${\cal L}_{\rho^{0}\pi\pi}$ is from ref.[1].
These vertices lead to the amplitude
\begin{eqnarray}
\lefteqn{{\cal M}_{K^{-*}\rightarrow K^{-}\pi^{+}\pi^{-}}=
{2\over f_{\pi}}\varepsilon^{\mu\nu\alpha\beta}\epsilon_{\mu}(p')
p'_{\nu}k_{+\alpha}k_{-\beta}\{{1\over 2\sqrt{2}\pi^{2}g}}\nonumber \\
&&({2\over f_{\pi}})^{2}(1-{6c\over g}+{6c^{2}\over g^{2}})
+\frac{N_{c}}{\pi^{2}g^{3}}[\frac{1}{(p'-k_{-})^{2}-m^{2}_{K^{*}}}
-\frac{1}{(p'-p)^{2}-m^{2}_{\rho}}]\},
\end{eqnarray}
where $p'$, p, $k_{-}$ are momentum of $\eta'$, $\eta$, and $\pi^{-}$
respectively.
The width calculated is
\begin{equation}
\Gamma(K^{-*}\rightarrow K^{-}\pi^{+}\pi^{-})=1.21keV.
\end{equation}
In the same way, the amplitude of $K^{-*}\rightarrow \bar{K}^{0*}
\pi^{-}\pi^{0}$ has been obtained
\begin{eqnarray}
\lefteqn{{\cal M}_{K^{-*}\rightarrow \bar{K}^{0}\pi^{-}\pi^{0}}=
{2\over f_{\pi}}\varepsilon^{\mu\nu\alpha\beta}\epsilon_{\mu}(p')
p'_{\nu}k_{0\alpha}k_{-\beta}\{{1\over 2\pi^{2}g}
({2\over f_{\pi}})^{2}(1-{6c\over g}+{6c^{2}\over g^{2}})}
\nonumber \\
&&-\frac{N_{c}}{\sqrt{2}\pi^{2}g^{3}}[\frac{1}
{(p'-k_{0})^{2}-m^{2}_{K^{*}}}+\frac{1}
{(p'-k_{-})^{2}-m^{2}_{K^{*}}}]
+\frac{\sqrt{2}N_{c}}{\pi^{2}g^{3}}
\frac{1}{(p'-p)^{2}-m^{2}_{\rho}}\}.
\end{eqnarray}
the width calculated is
\begin{equation}
\Gamma(K^{-*}\rightarrow \bar{K}^{0}\pi^{0}\pi^{-})=1.23keV.
\end{equation}
The total width is 2.65 keV which is below the limit. From eqs.(59,61)
it can be seen that there are cancellations too
in these two amplitudes.
In these processes there are subprocesses of normal parity and abnormal
parity and the relative signs between these subprocesses
have been determined without any ambiguity. Because all the vertices
are revealed from the lagrangian(1). This is the universality of the
couplings of this theory. Both
the smallness of the phase space
and the cancellations cause the smallness of the branch ration of
$K^{*}\rightarrow K\pi\pi$.

{\large\bf Electromagnetic decays of mesons}\\
In this section the processes: $\phi\rightarrow \eta\gamma$,
$\eta\rightarrow \gamma\gamma$, $\eta'\rightarrow \rho\gamma, \omega
\gamma$, and $\eta'\rightarrow \gamma\gamma$ have been studied.
These processes have been studied in ref.[7]. The formulas
obtained in this theory are the same with the ones derived from
gauging Wess-Zumino lagrangian in ref.[7]. However, as mentioned
above, in this theory there is universality of couplings and VMD
is not an input.
In the vertices of these processes the number of $\gamma_{5}$ is odd
and they are processes of abnormal parity. In ref.[1], $<\bar{\psi}
\gamma_{5}\psi>$ has been evaluated(see eq.(177) of ref.[1]). In the
same way $<\bar{\psi}\lambda_{8}\gamma_{5}\psi>$ has been evaluated.
We are interested in the vertices of $\eta vv$ and $\eta' vv$ which
have been found as
\begin{eqnarray}
\lefteqn{{\cal L}_{\eta vv}=
\frac{N_{c}}{(4\pi)^{2}}{4\over g^{2}}\varepsilon
^{\mu\nu\alpha\beta}\eta\{(-\sqrt{{2\over 3}}sin\theta+
{1\over \sqrt{3}}cos\theta)(\partial_{\mu}\omega_{\nu}\partial_
{\alpha}\omega_{\beta}+\partial_{\mu}\rho^{i}_{\nu}\partial_{\alpha}
\rho^{i}_{\beta})}\nonumber \\
&&-(\sqrt{{2\over 3}}sin\theta+{2\over \sqrt{3}}cos\theta)
\partial_{\mu}\phi_{\nu}\partial_{\alpha}\phi_{\beta}\},\nonumber \\
&&{\cal L}_{\eta' vv}=\frac{N_{c}}{(4\pi)^{2}}{4\over g^{2}}\varepsilon
^{\mu\nu\alpha\beta}\eta'\{(\sqrt{{2\over 3}}cos\theta+
{1\over \sqrt{3}}sin\theta)(\partial_{\mu}\omega_{\nu}\partial_
{\alpha}\omega_{\beta}+\partial_{\mu}\rho^{i}_{\nu}\partial_{\alpha}
\rho^{i}_{\beta})\nonumber \\
&&+(\sqrt{{2\over 3}}cos\theta-{2\over \sqrt{3}}sin\theta)
\partial_{\mu}\phi_{\nu}\partial_{\alpha}\phi_{\beta}\},
\end{eqnarray}
where $\theta$ is the mixing angle between $\eta$ and $\eta'$.
Combining VMD(eqs.(16,17)) and eqs.(63), the decay widths of the
physical processes have been found
\begin{eqnarray}
\lefteqn{\Gamma(\eta\rightarrow \gamma\gamma)=\frac{\alpha^{2}}
{16\pi^{3}}\frac{m^{3}_{\eta}}{f^{2}_{\eta}}(2\sqrt{2\over 3}
sin\theta-{1\over \sqrt{3}}cos\theta)^{2}),}\nonumber \\
&&\Gamma(\eta'\rightarrow \gamma\gamma)=\frac{\alpha^{2}}
{16\pi^{3}}\frac{m^{3}_{\eta'}}{f^{2}_{\eta'}}(2\sqrt{2\over 3}
cos\theta+{1\over \sqrt{3}}sin\theta)^{2},\nonumber \\
&&\Gamma(\phi\rightarrow\eta\gamma)=\frac{\alpha}{48\pi^{4}g^{2}}
\frac{m^{3}_{\phi}}{f^{2}_{\eta}}(1-{m^{2}_{\eta}\over m^{2}_{\phi}})
^{3}(\sqrt{2\over 3}sin\theta+{2\over \sqrt{3}}cos\theta)^{2},
\nonumber \\
&&\Gamma(\omega\rightarrow \eta\gamma)=\frac{\alpha}{96\pi^{4}
g^{2}}\frac{m^{3}_{\omega}}{f^{2}_{\eta}}(1-\frac{m^{2}_{\eta}}
{m^{2}_{\omega}})^{3}
(-\sqrt{{2\over 3}}sin\theta+{1\over \sqrt{3}}
cos\theta)^{2},\nonumber \\
&&\Gamma(\rho\rightarrow\eta\gamma)=\frac{3\alpha}{32\pi^{4}g^{2}}
\frac{m^{3}_{\rho}}{f^{2}_{\eta}}(1-\frac{m^{2}_{\eta}}
{m^{2}_{\rho}})^{3}
(-\sqrt{{2\over 3}}sin\theta+{1\over \sqrt{3}}
cos\theta)^{2},\nonumber \\
&&\Gamma(\eta'\rightarrow\rho\gamma)=\frac{9\alpha}{32\pi^{4}g^{2}}
\frac{m^{3}_{\eta'}}{f^{2}_{\eta'}}
(1-\frac{m^{2}_{\rho}}{m^{2}_{\eta'}})^{3}
(\sqrt{{2\over 3}}cos\theta+{1\over \sqrt{3}}sin\theta)^{2},
\nonumber \\
&&\Gamma(\eta'\rightarrow\omega\gamma)=\frac{\alpha}{32\pi^{4}g^{2}}
\frac{m^{3}_{\eta'}}{f^{2}_{\eta'}}
(1-\frac{m^{2}_{\omega}}{m^{2}_{\eta'}})^{3}
(\sqrt{{2\over 3}}cos\theta+{1\over \sqrt{3}}sin\theta)^{2}.
\end{eqnarray}
There are two values for the mixing angle $\theta$[4]. \(\theta=
-10^{0}\) from quadratic mass formula and \(\theta=-23^{0}\)
from linear mass formula. According to ref.[8],
the two photon decays of $\eta$ and $\eta'$ favor \(\theta
=-20^{0}\). In this theory, \(\theta=-20^{0}\) gives a better fits
too.
In chiral limit, we take \(f_{\eta}=f_{\eta'}=f_{\pi}\).
The numerical results are shown in Table I.

{\large\bf Conclusion}\\
In this paper two new mass formulas have been obtained. The theoretical
values of the hadronic decay rates are lower than data. The worse one
is $\phi\rightarrow K\bar{K}$ which is less than data by $30\%$.
The corrections from strange quark mass should take responsibility
for these deviations. In ref.[5] the corrections of strange quark
mass to $f_{K}$ and $f_{\eta}$ have been studied. All
other results agree with data well. Especially, this theory provides
better understanding of the smallness of $\Gamma(K_{1}\rightarrow K
\rho, K\omega)$, $\Gamma(K^{*}\rightarrow K\pi\pi)$ and the decay of
$\eta'\rightarrow \eta\pi\pi$. $f_{\pi}$, $m_{\pi}$, $m_{\eta}$,
$m_{\rho}$, and g are not only inputs here and they are also inputs of
ref.[1]. It is needed to point out that the introduction of
vector and axial-vector fields to the theory is not based on
gauge invariance, but on minimum coupling principle. This opens a door
to introduce other mesons to the theory. In chiral limit, the cut-off
determined in ref.[1] is 1.6 GeV. The mass of $f_{1}(1510)$ is closer
to this value. However, we still obtain pretty good result of the
decay $f_{1}(1510)\rightarrow K^{*}K$.
\begin{table}[h]
\begin{center}
\caption {Table I Summary of the results}
\begin{tabular}{|c|c|c|} \hline
    &  Experimental  &  Theoretical  \\ \hline
$f_{\pi}$   & 0.186GeV    & input         \\ \hline
$m_{\pi}$   & 0.138 Gev    & input         \\ \hline
$m_{K^{+}}$     & 0.494 Gev    & input         \\ \hline
$m_{K^{0}}$     & 0.498 Gev    & input         \\ \hline
$m_{\eta}$   & 0.548 Gev    & input         \\ \hline
$m_{\eta'}$   & 0.958 Gev    & input         \\ \hline
$m_{\rho}$  & 0.77GeV     & input         \\ \hline
$m_{K^{*}}$  & 0.892GeV    & input         \\ \hline
$m_{\phi}$  & 1.02GeV    & input         \\ \hline
g         &              & 0.35  input   \\ \hline
$m_{K_{1}}$     & 1402$\pm$7MeV    & 1.51 GeV     \\ \hline
$m_{f_{1}(1510)}$     & 1512$\pm$4MeV    & 1.64 GeV     \\ \hline
$g_{\phi\gamma}$ & $0.081(1\pm 0.05)$ $GeV^{2}$
& 0.086 $GeV^{2}$ \\ \hline
$<r^{2}>_{K}$ & $0.34\pm0.05fm^{2}$ & $0.33 fm^{2}$     \\ \hline
$\lambda_{+}(K^{+}_{l3})$ & $0.0286\pm 0.0022$&0.0239
 \\ \hline
$\xi(K^{+}_{l3})$ &$-0.35\pm 0.15$ & -0.284\\  \hline
\end{tabular}
\end{center}
\end{table}
\begin{table}
\begin{center}
\begin{tabular}{|c|c|c|} \hline
$\lambda_{+}(K^{0}_{l3})$ & $0.03\pm 0.0016$&0.0245
 \\ \hline
$\xi(K^{0}_{l3})$ &$-0.11\pm 0.09$ & -0.287\\  \hline
$\Gamma(K^{+}_{e3})$&$0.256(1\pm 0.015)\times 10^{-17}$GeV&$0.233
\times 10^{-17}$GeV\\  \hline
$\Gamma(K^{0}_{e3})$&$0.493(1\pm 0.016)\times 10^{-17}$GeV&$0.483
\times 10^{-17}$GeV\\  \hline
$B(\tau\rightarrow K^{*}(892)\nu)$ &$(1.45\pm 0.18)\%$
 &$1.46\%$   \\ \hline
$\Gamma(\tau\rightarrow K_{1}(1400)\nu)$ & &$0.373\%$ \\  \hline
$\Gamma(\phi\rightarrow K^{0}\bar{K}^{0})$ & $1.52(1\pm 0.03)$ MeV
& 1.11MeV \\ \hline
$\Gamma(\phi\rightarrow K^{+}K^{-})$ &$2.18(1\pm 0.03)$ & 1.7MeV \\
\hline
$K^{*}(892)\rightarrow K\pi)$ & $49.8\pm 0.8$MeV &39.4 M3V \\
\hline
$\Gamma(K^{+*}\rightarrow K^{+}\gamma)$&$50.3(1\pm 0.11)$keV&43.5keV\\
\hline
$\Gamma(K^{0*}\rightarrow K^{0}\gamma)$&$116.2(1\pm 0.10)$keV
&175.4keV\\ \hline
$f_{1}(1510)\rightarrow K^{*}(892)\bar{K})$ & $35\pm 15 $MeV
&22.MeV \\ \hline
$\Gamma(K_{1}(1400)\rightarrow K^{*}(892)\pi)$ & $163.6(1\pm 0.14)$ MeV
 &126 MeV\\  \hline
$B(K_{1}(1400)\rightarrow K\rho)$ & $(3.0\pm 3.0)\%$
& $11.1\%$ \\ \hline
$B(K_{1}(1400)\rightarrow K\omega)$ & $(2.0\pm 2.0)\%$&
$2.4\%$ \\  \hline
$\Gamma(K_{1}\rightarrow K\gamma)$ &  &440keV\\ \hline
$\Gamma(\eta'\rightarrow\eta\pi^{+}\pi^{-})$ &$87.8(\pm 0.12)$keV
&85.7keV\\ \hline
$\Gamma(\eta'\rightarrow\eta\pi^{0}\pi^{0})$ &$41.8(\pm 0.11)$keV
&48.6keV\\ \hline
\end{tabular}
\end{center}
\end{table}
\begin{table}
\begin{center}
\begin{tabular}{|c|c|c|} \hline
$\Gamma(\eta\rightarrow\gamma\gamma)$ &$0.466(1\pm 0.11)$keV
&0.619keV\\ \hline
$\Gamma(\phi\rightarrow\eta\gamma)$ & $56.7(1\pm 0.06)keV$& 91.4keV
  \\ \hline
$\Gamma(\rho\rightarrow\eta\gamma)$ &$57.5(1\pm 0.19)$
&61.4keV\\ \hline
$\Gamma(\omega\rightarrow\eta\gamma)$ &$7.0(1\pm 0.26)keV
$ &7.84keV \\ \hline
$\Gamma(\eta'\rightarrow\gamma\gamma)$ &$4.26(1\pm 0.14)$keV
&4.88keV\\ \hline
$\Gamma(\eta'\rightarrow\rho\gamma)$ &$60.7(1\pm 0.12)$keV
&63.0keV\\ \hline
$\Gamma(\eta'\rightarrow\omega\gamma)$ &$6.07(1\pm 0.18)$keV
 &5.86keV \\ \hline
\end{tabular}
\end{center}
\end{table}

The author likes to thank C.S.Lam, K.F.Liu, and M.L.Yan
for discussion.
This research is partially supported by DE-91ER75661.

\end{document}